# STATUS OF THE MINIMAL SUPERSYMMETRIC STANDARD MODEL [*][†]

Stefan Pokorski [‡]

*Max-Planck-Institute fuer Physik, Werner - Heisenberg - Institute*
*Foehringer Ring 6, 80805 Munich, Germany*

**Abstract**

Recent results in study of the Minimal Supersymmetric Standard Model as the effective low energy theory give important hints for experimental search for supersymmetry. Also, in the bottom - up approach to explore weak scale - GUT scale connection, they constrain strongly physics at the GUT scale. Several theoretical ideas about the GUT physics are in stunning agreement with the low energy data.

[*]This work was supported in part by the Polish Committee for Scientific Research and by European Union under contract CHRX-CT92-0034.
[†]Invited talk given at SUSY95, l'École Polytechnique, Palaiseau, France, May 15–19, 1995
[‡]On leave from the Institute of Theoretical Physics, Warsaw University

# 1 INTRODUCTION

Soon after the Wess-Zumino paper [1], attempts have been made to find realistic applications of supersymmetry to particle physics [2]. Those models use the mechanism of spontaneous breaking of supersymmetry and it turned out that this was the main obstacle for constructing a realistic extension of the standard electroweak theory [3]. Therefore, attention has been moved towards explicit but soft breaking of supersymmetry [4, 5, 6] and in 1981 a phenomenological model has been proposed [4] which forms the basis of what is now called the Minimal Supersymmetric Standard Model (MSSM). It has been paralleled by a systematic study of soft global supersymmetry breaking [6] and soon appeared very important papers on the possibility of supergravitational origin of the soft mass terms [7].

The main two aspects of the MSSM are:

**(I)** MSSM as the effective low energy theory which addresses the "hierachy problem"[8] ( the soft supersymmetry breaking scale is a physical cut-off to the Standard Model (SM)) and provides phenomenologically viable extension of the SM. The three underlying assumptions which define the model are:

(a) Minimal particle content consistent with the observed particle spectrum and with supersymmetry;

(b) R-parity conservation;

(c) Most general soft supersymmetry breaking terms consistent with the SM gauge group and with constraints from the observed suppresion of FCNC.

In addition to those assumptions the inherent part of the MSSM is the expectation that the superpartner masses are $< \mathcal{O}(1 \text{ TeV})$; otherwise the original motivation for the MSSM disappears.

**(II)** MSSM as the framework to study weak scale - GUT (string) scale connection.

This aspect of the model has been strongly emphasized in the early papers [9, 4, 5, 10]. Here the additional basic assumption is that the MSSM is the correct theory up to the GUT (string) scale [11]. That means, in particular, that physics remains perturbative up to those scales (but, possibly, with non-perturbative regime around the Planck scale). One way to explore the weak scale - GUT scale connection is the bottom - up phenomenological approach in which we use the available experimental information together with those few general assumptions to search for the underlying theory at very large scales. Within the MSSM, weak scale - GUT scale connection can be studied in a fully quantitative way, with high scale parameters unambigously connected to low energy observables. One should also stress that weak scale - GUT scale connection is a very natural (by easing the hierachy problem supersymmetry paves the road to Grand Unification) and most desirable (large number of free parametrs - many soft terms - should have its explanation in very high energy physics) aspect of the MSSM. This approach is certainly a useful supplement (and constraint!) to the top down one.

Recent results in study of the Minimal Supersymmetric Standard Model as the effective low energy theory give important hints for experimental search for supersymmetry. Also, in the bottom - up approach to explore weak scale - GUT scale connection, they constrain strongly physics at the GUT scale. Several theoretical ideas about the GUT physics are in stunning agreement with the low energy data.

# 2 MSSM AS THE LOW ENERGY EFFECTIVE THEORY

The outstanding low energy signatures of the MSSM is the Higgs sector and the superpartner spectrum. Not only supersymmetry provides a rationale for the Higgs mechanism but it is also well known that the mass of the lightest Higgs boson is strongly constrained (due to supersymmetry and the minimal particle content there is no independent Higgs boson self-coupling in the model; Higgs boson self-interaction originates only from the D-terms and it is fixed in terms of the gauge couplings)

$$M_h^2 < M_Z^2 \cos^2 2\beta + \frac{12g^2}{16\pi^2} \frac{m_t^4}{M_W^2} \log \frac{M_{\tilde{t}_1}^2 M_{\tilde{t}_2}^2}{m_t^4} \qquad (1)$$

Very important one-loop corrections [12] (the second term on the rhs of eq.(1)) increase the bound significantly above the tree level bound (for heavy top quark and $M_{SUSY} \sim \mathcal{O}(1 \text{ TeV})$) but do not change qualitative conclusions. The non-renormalization theorem assures that the corrections are finite and depend only logarithmically on the soft supersymmetry breaking scale. Accurate numerical estimates for the supersymmetric Higgs boson mass (important for the planned experiments) as a function of $m_t$, $M_{SUSY}$ and the L-R mixing include now also two-loop corrections [13] which partially cancel the large one-loop effect.

It is instructive to look at the SM as a heavy superpartner limit of the MSSM in which sparticles decouple from the electroweak observables (in practice, this occurs for masses > (300 − 400) GeV). What do we learn about the Higgs sector then?

The SM fits to the precision electroweak data [14, 15, 16, 17, 18, 19] send us two important messages. First, they confirm the electroweak symmetry breaking by the Higgs mechanism with few per mille accuracy! Second, they give some evidence (although not yet fully satisfactory from the statistical point of view) for a light Higgs boson. The usual strategy is as follows: in terms of the best measured observables $G_F$, $\alpha_{EM}$ and $M_Z$ one calculates the observables $M_W$, all partial widths of $Z^0$ and all asymmetries at the $Z^0$ pole. The value of the top quark mass from the Fermilab measurements can also be included in the fit. The SM calculation of the electroweak observables is performed with one loop accuracy and with the leading higher order effects also included. The parameters $m_t$, $M_h$ and $\alpha_s(M_Z)$ are determined by a fit. The best fit to the spring 95 LEP data [18], the SLAC result for the leptonic effective Weinberg angle [20], $M_W = 80.33 \pm 0.17$ GeV [21], $m_t = (181 \pm 12)$ GeV [22] and with the up-dated value of $\Delta\alpha_{EM}^{hadr} = 0.0280 \pm 0.0007$ [23] gives [24]:[§] (for the sake of later discussion in version B we include in the fit the low energy measurement of $\alpha_s$ [26]: $\alpha_s(M_Z) = 0.112 \pm 0.005$).

**Table 1.** Results of a fit in the SM. All masses in GeV.

| fit | $m_t$ | $\Delta m_t$ | $M_h$ | $\Delta M_h$ | $\alpha_s(M_Z)$ | $\Delta\alpha_s(M_Z)$ | $\chi^2$ | d.o.f |
|-----|-------|--------------|-------|--------------|-----------------|----------------------|----------|-------|
| A   | 171.3 | $^{+11.5}_{-9.7}$ | 66 | $^{+117}_{-45}$ | 0.123 | 0.005 | 12.6 | 12 |
| B   | 172.0 | $^{+10.5}_{-9.3}$ | 59 | $^{+96}_{-37}$ | 0.120 | 0.005 | 15.5 | 13 |

We can interpret the SM fits as the MSSM fits with all superpartners heavy enough to be decoupled. Supersymmetry then just provides a rationale for a light Higgs boson: $M_h \sim \mathcal{O}(100$ GeV). We can, therefore, expect that the MSSM with heavy enough superpartners gives as good a fit to the precision electroweak data as the SM, with $m_t \sim 170-180$ GeV (depending slightly on the value of $\tan\beta$). This is seen in Fig.1b where we show the $\chi^2$ values in the MSSM with the proper dependence of $M_h$ on $m_t$, $\tan\beta$ and SUSY parameters included, with fitted $m_t$ and $\alpha_s(M_Z)$ and with all SUSY mass parameters fixed at 500 GeV. The $\chi^2$ values in the minima are very closed to the SM three parameter ($m_t$, $M_h$, $\alpha_s(M_Z)$) fit.

---

[§]Those results agree very well with another recent fit [25].

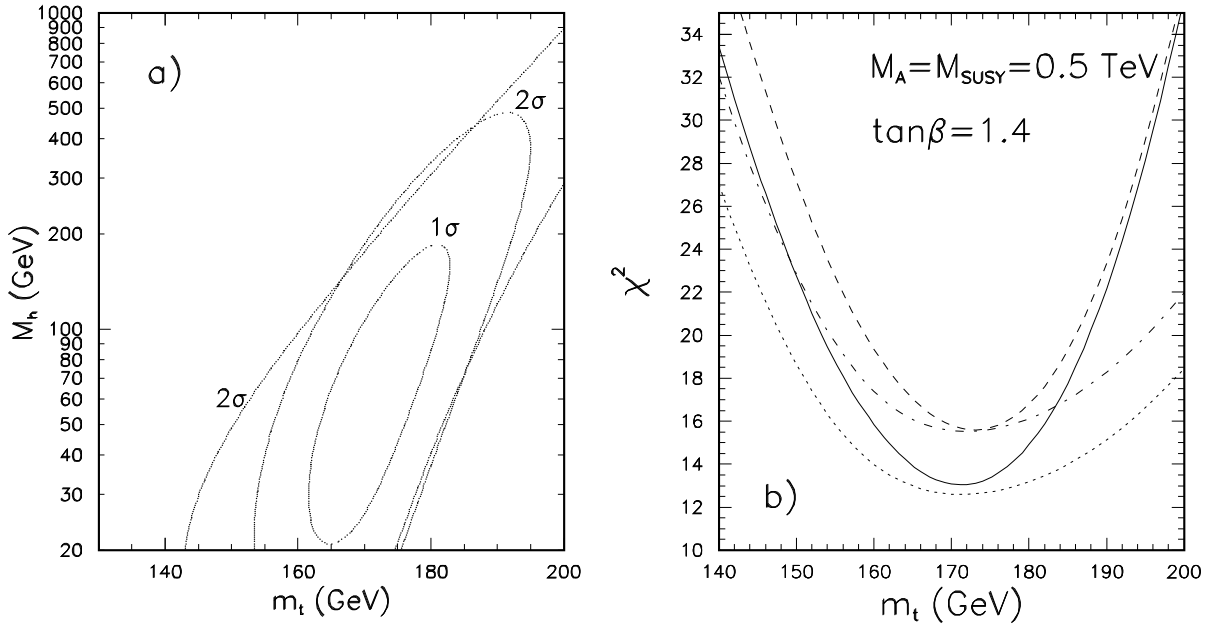

**Figure 1:** a) 1$\sigma$ and 2$\sigma$ contours in the SM (open lines are for a fit without $R_b$ and $m_t$ included); b) $\chi^2$ in the MSSM with heavy superpartners (solid and dashed lines) compared with the SM (dotted and dashed-dotted lines); upper curves are always for version B.

Although the SM fit and the MSSM fit with heavy superpartners are globally good, it is worth observing that about 70% of the contribution to the overall $\chi^2$ comes from the single variable $R_b = \Gamma_{Z \to b\bar{b}}/\Gamma_h$ which remains almost 3$\sigma$ higher than the theoretical prediction. This deviation of the measured $R_b$ from the SM predition (i.e. of $\Gamma_{Z \to b\bar{b}}$, as the total hadronic width $\Gamma_h$ is independently determined from other LEP measurements) has one very interesting aspect when correlated with the fitted value of $\alpha_s$ [27, 17, 28]. The SM fits give $\alpha_s = 0.123 \pm 0.005$ and this result essentially follows from the fit to the experimental value of $\Gamma_h$ where

$$\Gamma_h^{Th} = \Gamma_{h,EW}^{Th}(1 + \frac{\alpha_s(M_Z^2)}{\pi} + ...) \qquad (2)$$

(the QCD corrections and the electroweak contributions to $\Gamma_h$ factorize to a good approximation). For the SM value of $\Gamma_{h,EM}^{Th}$ one needs $\alpha_s(M_Z) \sim 0.123$ to fit the measured value of $\Gamma_h$. It is somewhat larger then the value obtained from low energy data [26]. ¶ (It is interesting to repeat the SM fit with the low energy measurement [26] $\alpha_s(M_Z) = 0.112 \pm 0.005$ included in the fit. Those results are also shown in in Table 1 (case B). The parameters of the fit remain unaltered but the overall $\chi^2$ is larger by $\sim 3$.) If however, we assume some new physics in $\Gamma_{Z \to b\bar{b}}$ and perform the SM fit to the difference $\Gamma_h - \Gamma_{Z \to b\bar{b}}$ then the result for the central value of $\alpha_s$ reads $\alpha_s(M_Z) \sim 0.11$! in much better agreement with the low energy results. $\|$

---

¶One can argue that the determination of $\alpha_s(M_Z)$ based on the deep inelastic (Euclidean) analysis is more precise than from the experiments in the Minkowskean region (jet physics, $\tau$ decays). Low value of $\alpha_s(M_Z)$ is also consistent with lattice calculation and has some theoretical support (for review of all those points see M.Shifman, ref.[28]). So, with proper attention to the unsettled controversy and to the fact that jet physics and $\tau$ decays give larger values, we are going to explore the assumption that the low energy determination of $\alpha_s(M_Z)$ is the correct one.

Thus it is conceivable that the measurement of $R_b$ is not just a statistical fluctuation but an evidence for new physics and it is very interesting to perform a global fit to the electroweak observables in the MSSM with supersymmetric masses kept as free parameters. In particular we can ask the following two questions [30, 31, 24]:
a) can we improve $R_b$ without destroying the excellent fit to the other observables?
b) if we achieve this goal, what are the predictions for sparticle masses?

Precision tests of the MSSM have been discussed by several groups [33, 34, 15, 35, 36, 32]. In particular, first global fit to the electroweak data within the MSSM parametrized in terms of few parameters at the GUT scale is given in ref. [34]. In ref. [35] the so called $\epsilon_i$ parametrization is used and the rôle of light superpartners is studied in some detail. Here we present the results of a global fit to the electroweak observables with, for the first time, all the (relevant) low energy parameters of the MSSM treated as independent variables in the fit [30, 24]. The set of low energy parameters suggested by the precision data may give interesting hints on physics at the GUT scale.

The strategy is analogous to the one used for the SM and the calculation is performed in the on–shell renormalization scheme [37] and with the same precision as the analogous calculation in the SM, i.e. with all supersymmetric oblique and process dependent one–loop corrections [38, 39], and also the leading higher order effects included.

Although the MSSM contains many free SUSY parameters several of them are irrelevant. In ref.[24] the fitted parameters are: $m_t$, $\alpha_s(M_Z)$, $\tan\beta$, $M_A$, $\mu$, $M_{g2}$ ($M_{g1} = (5/3)\tan^2\theta M_{g2}$; it is of little importance for the results of the fit but fixes the parameters of the lightest supersymmetric particle LSP), $M_{\tilde{b}_L}$, $M_{\tilde{b}_R}$, $M_{\tilde{t}_R}$, $A_b$, $A_t$ and $M_{\tilde{l}_L}$ (a common mass for all left handed sleptons). The gluino mass, the first two generation squark masses and the right handed slepton masses are irrelevant for the fit and are always kept heavy. For details we refer the reader to ref.[24] (see also [31] for low $\tan\beta$ and [49, 50] for large $\tan\beta$ results). The main point is that in the MSSM the electroweak observables exhibit certain "decoupling": all of them but $R_b$ are sensitive mainly to the left-handed slepton and squark masses and depend weakly on the right-handed sfermion masses and on the gaugino and Higgs sectors; on the contrary, $R_b$ depends strongly just on the latter set of variables [40, 41] and very weakly on the former. We can then indeed expect to increase the value of $R_b$ (particulary in the low and very large $\tan\beta$ region [42] without destroying the perfect fit of the SM to the other observables[30, 31, 24]. However, chargino, right-handed stop and charged Higgs boson masses are also crucial variables for the decay $b \to s\gamma$ and this constraint has to be included [43, 44, 24]. In general, one obtains acceptable $BR(b \to s\gamma)$ due to cancellations between $W^+$, $H^+$, $\chi^+$ and $\tilde{t}_R$ loops.

---

$\|$ New electroweak data have been presented at the International Europhysics Conference on High Energy Physics (Brussels, 27 July – 2 August, 1995). The main change are the values of $R_b$ and $R_c$: $R_b = 0.2219 \pm 0.0017$, $R_c = 0.1543 \pm 0.0074$. Since identification of the $c$ quarks is more difficult than of the $b$ quarks, the experimental groups also present the value of $R_b = 0.2206 \pm 0.0016$ obtained under the assumption that $R_c$ is fixed to its SM value $R_c = 0.172$. The results discussed in this talk remain unchanged if we adopt the latter value of $R_b$ and disregard the new value of $R_c$ as unreliable. The MSSM cannot explain any significant departure of $R_c$ from the SM prediction [29].

**Table 2.** Results of a fit in the MSSM.

| fit | $\tan\beta$ | $m_t$ | $\alpha_s(M_Z)$ | $\chi^2$ | $R_b$ |
|---|---|---|---|---|---|
| A | IR | $178^{+5}_{-8}$ | $0.116^{+0.006}_{-0.004}$ | 10.3 | 0.218 |
| B | IR | $177^{+4}_{-6}$ | $0.114^{+0.004}_{-0.003}$ | 10.6 | 0.218 |
| A | $m_t/m_b$ | $172^{+8}_{-7}$ | $0.114 \pm 0.005$ | 10.2 | 0.219 |
| B | $m_t/m_b$ | $174^{+6}_{-7.3}$ | $0.113 \pm 0.004$ | 10.2 | 0.219 |

Let us present some quantitative results[24]. The best fit is obtained in two regions of very low (close to the quasi–IR fixed point for a given top quark mass, see e.g. [45] and references therein) and very large ($\sim m_t/m_b$) $\tan\beta$ values ( for early discussion of large $\tan\beta$ region see [47]) and is summarized in Table 2 (versions A and B are, as before, without and with the low energy value $\alpha_s(M_Z) = 0.112 \pm 0.005$ in the fit; the quality of the fit is similar in both cases as the MSSM fit to the electroweak data alone (case A) **gives** low value of $\alpha_s$). We recall (see Fig.1b) that in the fit with all superpartners heavy the best $\chi^2$ values read: $\chi^2 = 13(16)$, for $\tan\beta = 1.4$ and $\chi^2 = 13.3(16)$, for $\tan\beta = 50$ for fits without (with) low energy value for $\alpha_s$ included.

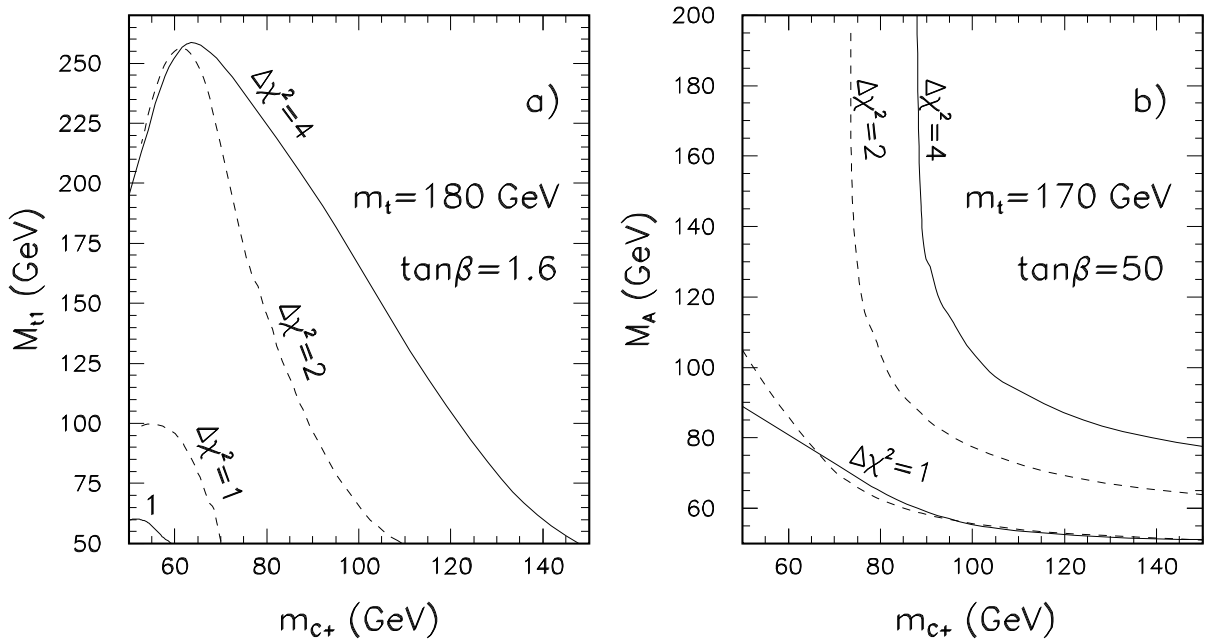

**Figure 2:** Upper bounds on the sparticle masses.

Increase of $\Gamma_{Z \to \bar{b}b}$ requires light stop and chargino (for low values of $\tan\beta$) or light $CP-$odd scalar and/or chargino and stop for large $\tan\beta$ and it is bounded from above by the experimental lower limits on the masses of those particles. A light and dominantly right-handed stop is obtained for $M_{\tilde{t}_L} \gg M_{\tilde{t}_R}$ and large L-R mixing. The fits give upper bounds on the light stop, chargino and CP-odd Higgs boson masses. In version A, when $\alpha_s$ runs free and is fitted only to the electroweak data, the best fit is better than the corresponding fit with all superpartners heavy by only $\Delta\chi^2 \sim 3$ (but then $\alpha_s(M_Z) = 0.123$). So, we obtain strong upper bounds at $1\sigma$ level but no 95% C.L. limits. They are shown in Fig.2a and 2b (dotted lines) for the stop and chargino masses in the low $\tan\beta$ region and for the pseudoscalar and chargino masses in the large $\tan\beta$ region, respectively. Stronger bounds are obtained in version B of the fits, i.e. with the low energy value of $\alpha_s(M_Z)$ included in the fits. They are shown in the same Figures (solid lines). The strongest bounds are obtained when $\alpha_s(M_Z)$ is fixed to its best fit value. The dependence of the strength of the bounds on the way we treat $\alpha_s$ in our fits is quite obvious from the earlier discussion of the depth of the minima in $\chi^2$. Thus, relying on the low energy measurements of $\alpha_s$ one predicts at 95% C.L. those particles to be within the reach of LEP2.

Furthemore, the parameter space selected by the fit is interesting from the theoretical point of view. Low and large values of $\tan\beta$ are theoretically most appealing [45, 47]. The hierarchy $M_{\tilde{t}_L} > M_{\tilde{t}_R}$ which is necessary for a good fit in the low $\tan\beta$ region can be viewed as a natural effect of the top quark Yukawa coupling in the renormalization group running from the GUT scale. For a good fit in the large $\tan\beta$ region this hierarchy is less pronounced, in agreement with $Y_t \approx Y_b$. Finally, the hierarchy $\mu << M_{g2}$ (i.e. higgsino–like lightest neutralino and chargino) is inconsistent with the mechanism of radiative electroweak symmetry breaking and universal boundary conditions for the scalar masses at the GUT scale in the minimal supergravity model. However, it is predicted in models with certain pattern of non–universal boundary conditions [48]. In the next section we explore in more detail this weak scale GUT scale connection in the framework of the MSSM. We shall see that several theoretical ideas are in stunning agreement with those low energy results.

## 3 WEAK SCALE - GUT SCALE CONNECTION IN MSSM

Among the most important issues there are:
1. supersymmetric grand unification;
2. onset of non-perturbative physics (fixed point structures in the low energy theory);
3. theory of flavour:
    a) fermion masses;
    b) origin and pattern of soft supersymmetry
        breaking and its low energy implications.

### 3.1 GAUGE COUPLING UNIFICATION

The gauge coupling unification [51] within the MSSM has been widely publicized as a successful prediction of SUSY-GUTs [9, 4, 10, 52, 53, 54]. It is also often discussed in the context of stringy unification, with $M_{ST} \sim 4 \times 10^{17}$ GeV [56, 57, 55, 58, 59, 60, 61]. Recently, we witness some progress in the quantitave study of the running of the gauge couplings in the MSSM due to proper inclusion of all low energy effects, such as the best precision of the input parameters at the electroweak scale and the non-leading contribution from the superparnter thresholds.

The unification idea is predictive with respect to the behaviour of the $SU(3) \times SU(2) \times U(1)$ gauge couplings if physics at the GUT scale is described in terms of only two parameters: $\alpha_U$ and $M_U$ (minimal unification). Then we can predict e.g. $\alpha_s(M_Z)$ in terms of $\alpha_{EM}(M_Z)$ and $\sin^2\theta_W(M_Z)$. Here we mean the coupling constants defined in the SM which, we assume, is the correct *renormalizable* theory at the electroweak scale. The value of $\alpha_{EM}(M_Z)$ is obtained from the on-shell $\alpha_{EM}^{OS} = 1/137.0359895(61)$ via the RG running, with .01% uncertainty due to the continuous hadronic contribution to the photon propagator and with logarithmic dependence on the top quark mass. The most precise value of $\sin^2\theta_W(M_Z)$ in the SM is at present obtained from its calculation in terms of $G_\mu$, $M_Z$, $\alpha_{EM}$ and the top quark and Higgs boson masses (it is worth remembering that $\sin^2\theta_W(M_Z)$ and $\alpha_s(M_Z)$ are known with 0.1% and 10% accuracy, respectively).

More precisely, the prediction for the strong coupling constant in addition depends on the superpartner spectrum which will be, hopefully, known from experiment. For now, these are free parameters but, fortunately, the dependence of the prediction for $\alpha_s(M_Z)$ on the supersymmetric spectrum can be described to a very good approximation by a single effective parameter, $T_{SUSY}$ [53, 45]. Thus, we get

$$\alpha_s(M_Z) = f(G_\mu, M_Z, \alpha_{EM}, m_t, M_h, T_{SUSY}) \qquad (3)$$

where [45]

$$T_{SUSY} = |\mu| \left(\frac{m_{\tilde{W}}}{m_{\tilde{g}}}\right)^{\frac{3}{2}} \left(\frac{M_{\tilde{l}}}{M_{\tilde{q}}}\right)^{\frac{3}{19}} \left(\frac{M_{A^0}}{|\mu|}\right)^{\frac{3}{19}} \left(\frac{m_{\tilde{W}}}{|\mu|}\right)^{\frac{4}{19}} \qquad (4)$$

We observe that the effective scale $T_{SUSY}$ depends strongly on the values of $\mu$ and of the ratio $m_{\tilde{W}}$ to $m_{\tilde{g}}$ but very weakly on the values of the squarks and slepton masses. Two comments are in order here:

1. The effective parametrization of the supersymmetric thresholds in the equation for $\alpha_s(M_Z)$ in terms of $T_{SUSY}$ is exact for one-loop RGE; with two-loop equations there is some (very weak) dependence on the details of the spectrum through the dependence of $M_{GUT}$ on the spectrum.

2. So far, we have assumed the SM to be the correct effective theory at the electroweak scale. That is, we have neglected all non-renormalizable remnants of the MSSM (i.e. higher dimension operators supressed as $\mathcal{O}(M_Z/m_{SUSY})$) in the procedure of extracting $\sin^2\theta_W(M_Z)$ from the data. This approximation (Leading Logarithmic Threshold approximation) may easily be inadequate if some superpartner masses are of the order of $M_Z$[62]. Proper inclusion of those non-leading effects has been recently accomplished in a series of papers [63, 64, 65] (by working directly in the MSSM at the electroweak scale) and they can contribute up to 1% correction to the input value of $\sin^2\theta_W(M_Z)$. This is a very large correction from the point of view of the prediction for $\alpha_s$. The non-leading corrections to $\sin^2\theta_W$ are negative (this can be traced back to the additional sources of the custodial $SU(2)_V$ breaking in the contributions to the $\Delta\rho$ parameter) and, therefore, tend to increase the predicted value of $\alpha_s$. One should stress, however, that the magnitude of those effects strongly depends on the details of the spectrum. The corrections are large with light left-handed squarks and sleptons of the third generation (they give large additional $SU(2)_V$ breaking contribution to $\Delta\rho$) but almost absent (in spite of very small $T_{SUSY}$) for the spectrum obtained from the best fit to the electroweak data discussed in Section 2, i.e. with very light higgsino and the right-handed squarks and with $M_{\tilde{t}_L}$ significantly larger than $M_{\tilde{t}_R}$[63]. ** Of course, with non-logarithmic corrections included, $\alpha_s$ is no longer a function of the single parameter $T_{SUSY}$ which, however, remains a useful variable for the presentation of the results.

The predictions [63, 66] for the strong coupling constant in this minimal unification scenario are shown in Fig.3. The results of the leading logarithmic approximation (solid lines) for the superpartner thresholds (i.e. obtained with two-loop RGE and with successive decoupling of each superpartner at its threshold in the one-loop $\beta$-functions) are compared with the full calculation (markers), with complete one-loop threshold corrections included. For the purpose of illustration in Fig.3a a superpartner spectrum which is generic for the minimal supergravity model with radiative breaking and universal boundary conditions for the soft scalar masses at the GUT scale [67] is used and in Fig.3b - a spectrum from the best fit to the electroweak data (see Section 2).

---

**I am grateful to C.Wagner for an interesting discussion of this point.

From Fig.3 one can draw several conclusions: Firstly, small $T_{SUSY}$ gives large values of $\alpha_s$. For e.g. $T_{SUSY} < 300$ GeV, $\alpha_s$ is predicted to be above 0.123 (0.121) for $m_t = 180(160)$ GeV. For $\alpha_s < 0.113$ we need $T_{SUSY} > 1$ TeV. It should be stressed that in GUT's, we have the relation

$$\frac{m_{\tilde{W}}}{m_{\tilde{g}}} \sim \frac{\alpha_2(M_Z)}{\alpha_3(M_Z)} \quad and \quad T_{SUSY} \sim |\mu| \left(\frac{\alpha_2(M_Z)}{\alpha_3(M_Z)}\right)^{\frac{3}{2}} \sim \frac{1}{7}\mu \quad (5)$$

Thus, $T_{SUSY} \sim 300$ GeV (1 TeV) corresponds to $\mu \sim 2$ TeV (7 TeV). In models with radiative electroweak breaking the $\mu$ parameter is correlated, for proper Higgs potential, with the soft parameters which determine the sfermion masses and large $\mu$ implies squark masses of the same order of magnitude (but not vice versa!). Of course, it is easy to have large $T_{SUSY}$ with small $\mu$ if $M_{\tilde{W}}/M_{\tilde{g}} \gg 1$. However, it is difficult to imagine such a dramatic modification of relation (5), unless the motivation for the minimal unification of couplings in the MSSM disappears (at least as long as the SUSY breaking mechanism is gauge independent).

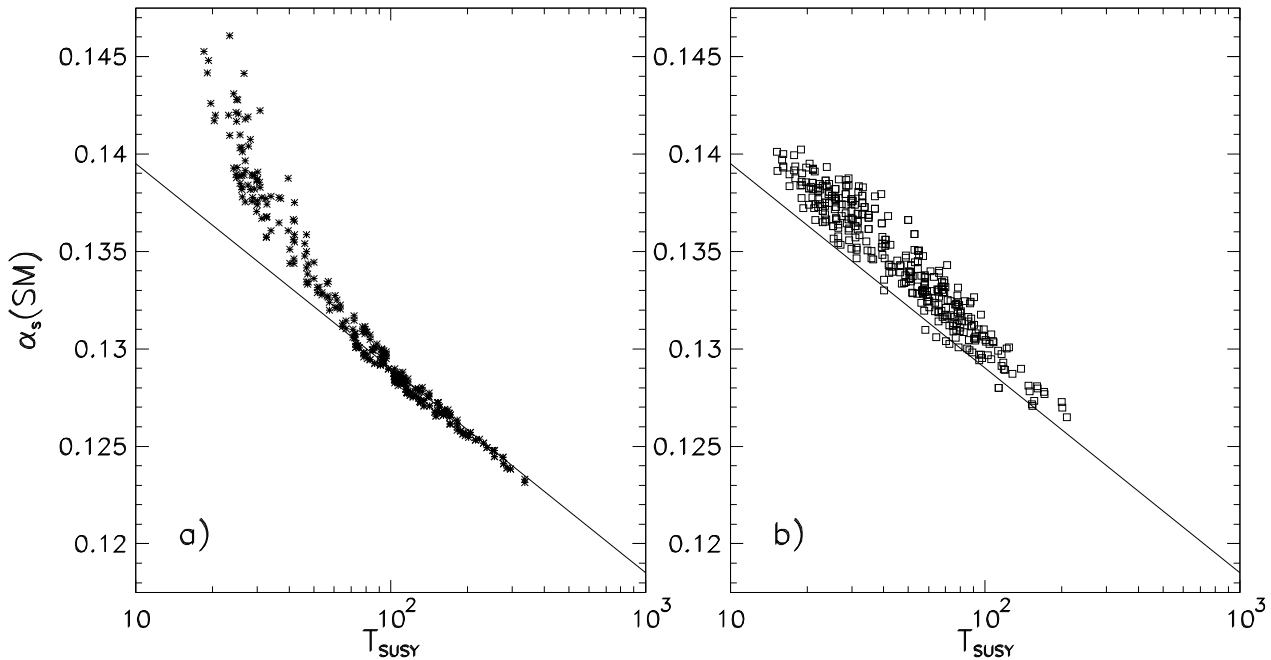

**Figure 3:** Predictions for $\alpha_s(M_Z)$ from minimal unification: ]a)for $m_t = 180$ GeV, $\tan\beta = 10$ with a generic spectrum from the minimal supergravity model, b)for $m_t = 180$ GeV, $\tan\beta = 1.6$ with the spectra from the best electroweak fits.

Secondly, the non-leading threshold corrections substantially increase the predicted value of $\alpha_s(M_Z)$ for $T_{SUSY} < 100$ GeV if the spectrum contains light left-handed sfermions, which is generically the case in models with radiative electroweak breaking and universal boundary conditions at the GUT scale. However, such corrections are small even for $T_{SUSY} \sim 20$ GeV with the spectra suggested by the best fit to the electroweak data and obtained in models with certain pattern of non-universal boundary conditions (see later).

The minimal unification may be too restrictive as it is generally expected that there are non-negligible GUT/string scale corrections to the running of the gauge couplings (such as heavy threshold and higher dimension operator effects). Then, strictly speaking, any predictivity is lost (at least highly model dependent). However, it is still very interesting to reverse the problem: take the values of all the three couplings at $M_Z$ as input parameters and use the bottom - up approach to study the convergence of the couplings in the framework of the MSSM. With the same precision calculation and as a function of the supersymmetric spectrum one can, then, discuss the mismatch of the couplings at any scale of interest and for any value of $\alpha_s(M_Z)$, within its 10% experimental uncertainty. It is convenient to introduce the "mismatch" parameters at the scale Q:

$$D_i(Q) = \frac{\alpha_i(Q) - \alpha_2(Q)}{\alpha_2(Q)} \qquad (6)$$

Of particular interest are $D_3(M_{GUT})$, where $M_{GUT}$ is defined as the scale of unification of the SU(2)xSU(1) couplings, and $D_i(M_{ST})$, $i = 1, 3$, where the string scale $M_{ST} = 4 \cdot 10^{17}$ GeV. Clearly, we get this way constraints on physics at the high scale, if it is supposed to have unification of couplings and the MSSM as the low energy effective theory.

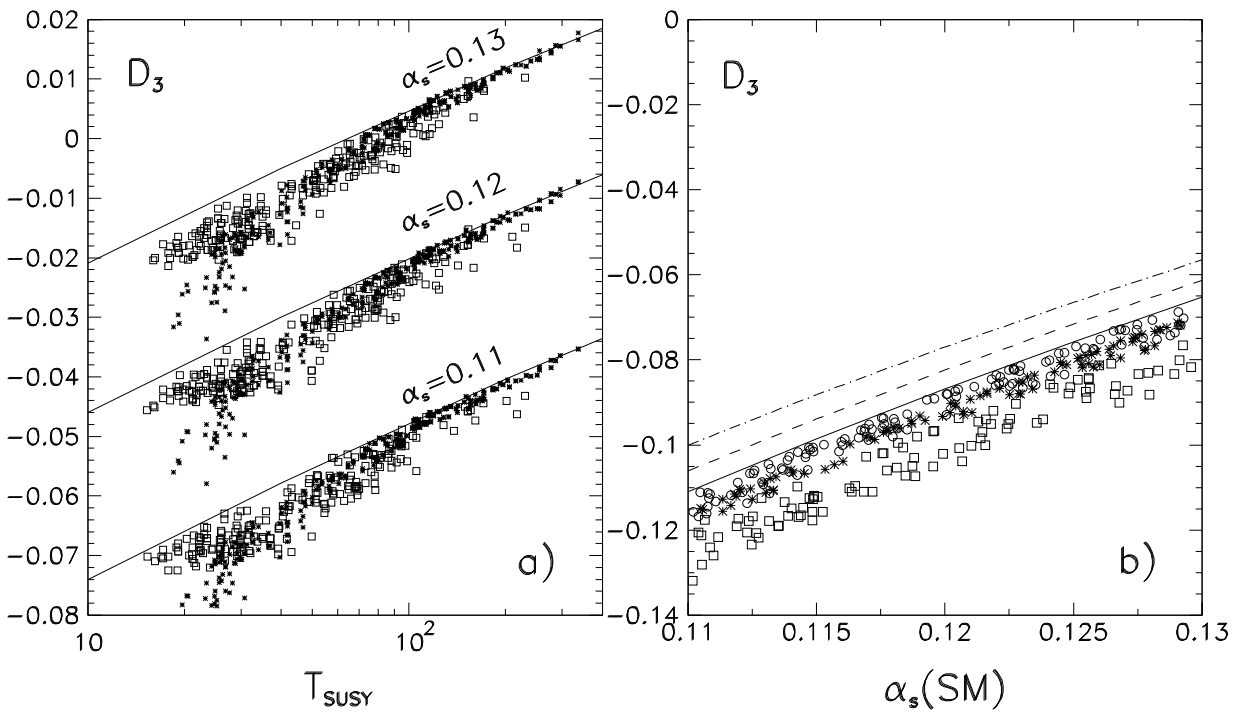

**Figure 4:** Mismatch of the couplings for $m_t = 180$ GeV a) at the unification scale defined by $\alpha_1(Q) = \alpha_2(Q)$ and as a function of $T_{SUSY}$; solid lines - leading logarithmic thresholds, squares and crosses - full calculation with the spectra as in Fig.3, b) at the string scale as a function of $\alpha_s$; straight lines - leading logarithmic results for $T_{SUSY} = 300$ GeV, 1 TeV, 5 TeV (bottom-up), markers-full calculation with different spectra.

We see from Fig.4 that in the bottom - up running the couplings do unify within a few percent accuracy even for low $\alpha_s(M_Z)$ and small values of $T_{SUSY}$. Is this mismatch a lot or a little depends on the GUT model and the expected magnitude of the GUT scale corrections in it [68, 69, 70, 71, 72, 65]. The mismatch of the couplings at the string scale is factor 2-3 larger (we assume here that the running of the couplings remains up to the string scale unaffected by new physics; this may not be true (see the next Section)) and it cannot be eliminated by any sensible superpartner spectrum. String unification requires, therefore, large string threshold corrections which conspire to give the effective unification scale $\sim 3 \cdot 10^{16}$. The scenario with $\alpha_1$ and $\alpha_2$ unified by treating the Kac-Moody level $k_1$ as a free parameter is not particularly helpful with regard to the coupling unification at $M_{ST}$ (and rather uneconomical).

### 3.2 INFRA – RED FIXED POINT STRUCTURE

The idea of Grand Unification is based on perturbative physics at least up to the unification scale $M_{GUT} \sim 2 \cdot 10^{16}$ GeV or even to the Planck scale $M_{Pl} \sim 2.4 \cdot 10^{18}$ GeV. There are, however, several reasons to expect that physics may become non-perturbative at the scale close to $M_{Pl}$. It is very interesting that this may have its low energy manifestation. A hint in this direction comes from the top quark mass which is stunningly close to its so-called quasi-infrared fixed point value. The Yukawa couplings may be determined in terms of a gauge coupling through the infra-red fixed point structure [73].

We recall the fixed point structure of the top quark Yukawa coupling $Y_t$ ($Y_t = h_t^2/4\pi$) in the MSSM. Ignoring the smaller electroweak couplings, $Y_t$ is related to the QCD coupling. We have ($t = \frac{1}{2\pi}\log\frac{M_{GUT}}{Q}$):

$$\frac{dY_t}{dt} = Y_t(b_t\alpha_s - c_tY_t), \qquad \alpha_s = \frac{g_3^2}{4\pi} \tag{7}$$

$$\frac{d\alpha_s}{dt} = -b_g\alpha_s^2 \tag{8}$$

with the fixed point solution for the ratio $Y_t/\alpha_s$ [73]:

$$Y_t^F(t) = \frac{b_t + b_g}{c_t}\alpha_s(t) \tag{9}$$

(in the MSSM $b_g = -3$, $b_t = 16/3$, $c_t = 6$).

One can also solve eqs.(7,8) explicitly [74]:

$$Y_t(t) = \frac{Y_t(0)E(t)}{1 + c_tY_t(0)F(t)} \tag{10}$$

with

$$E(t) = \left[\frac{\alpha_s(0)}{\alpha_s(t)}\right]^{\frac{b_t}{b_g}}, \qquad F(t) = \int_0^t E(t')dt' \tag{11}$$

It may happen that $Y_t^F(t)$ is not reached because of a too short a running but, nevertheless, $c_tY_t(0)F(t) >> 1$ and

$$Y_t(t) \approx Y_t^{QF}(t) = \frac{E(t)}{c_tF(t)} \tag{12}$$

i.e. the low energy value of $Y_t$ no longer depends on the initial value $Y_t(0)$. This is called the quasi-infra-red fixed point solution [75] and we have:

$$Y_t^{QF}(t) = Y_t^F(t)\frac{1}{1 - \left[\frac{\alpha_s(t)}{\alpha_s(0)}\right]^{1+\frac{b_t}{b_g}}} \tag{13}$$

This situation indeed occurs in the MSSM (for $b_g$, $b_t$, $c_t$ of the MSSM) for $Y_t(0) \sim \mathcal{O}(1)$ i.e. for the initial value still in the perturbative regime! Thus, not only $Y_t^{QF}(M_Z)$ is an upper bound for $Y_t(M_Z)$ but it can be reached at the limit of perturbative physics [76]. The quasi-infra-red fixed point prediction for the running top quark mass in the $\bar{M}S$ scheme, for $\alpha_s(M_Z)= 0.11$-$0.13$ and small and moderate values of $\tan\beta$ is approximately given by (see e.g. [36]):

$$m_t^{QF}(m_t) \approx 196 GeV[1 + 2(\alpha_s(M_Z) - 0.12)]\sin\beta \qquad (14)$$

The physical top quark mass (pole mass) is obtained by including QCD corrections which contribute $\mathcal{O}(10\ GeV)$ to the final result. Similar expression may also be found in the large $\tan\beta$ region in which both top and bottom Yukawa couplins are large.[††] In the numerical calculation two-loop RG equations are used. It is clear that the experimental value of $m_t \approx 180$ GeV is very close to its perturbative upper bound in the MSSM. To know how close, we need to know $\tan\beta$ but one is tempted to speculate that $m_t \approx m_t^{QF}$ and then it is easy to check that $\tan\beta$ must be either very small or very large. Large values of $\tan\beta \sim m_t/m_b$ have been independently discussed for some time as a solution to the $m_t/m_b$ hierachy [47, 81], particularly in the framework of SO(10) [47, 78]. It is also worth recalling that the low energy fits favour very low or very large values of $\tan\beta$, too (Section 2).

An interesting observation is that the GUT assumption about unification of the bottom and tau Yukawa couplings gives independent support to the idea that $Y_t(M_Z) = Y_t^{QF}(M_Z)$.[86, 45, 53, 46] Quantitatively, this conclusion depends on the values of $\alpha_s(M_Z)$ and $m_b$(pole) and on the threshold corrections to the relation $Y_b = Y_\tau$. Generically, however, strong interaction renormalization effects for $Y_b$ are too strong and large top quark Yukawa coupling contribution to the running of $Y_b$ is needed to balance them. For $Y_b = Y_\tau$ within 10%, $\alpha_s(M_Z) > 0.11$ and $m_b(pole) < 5$ GeV one gets $m_t \approx m_t^{QF}$ within 10%.[‡‡] One can conclude that the possibility of $m_t \approx m_t^{QF}$ is supported by several independent arguments (also models for dynamical determination of $Y_t$ give values close to the IR fixed point [87]).

So far we have been discussing the infra-red fixed point structure in the MSSM with perturbativity limit at $M_{GUT}$. If the MSSM is embedded in a unified model, we can also discuss physics at the scales from $M_{GUT}$ to $M_{Pl}$. Perturbativity limit should then be applied at $M_{Pl}$ and physics in the range $M_{GUT}$ to $M_{Pl}$ may have very interesting aspects, too [88]. It is quite conceivable that the unified model is not only not asymptotically free but that the gauge coupling has Landau pole at the Planck scale. Indeed, the $\beta$-function coefficient $b_G$ easily reaches large positive values in the presence of a few heavy matter representations. Thus, the intriguing possibility opens that physics at the GUT scale is determined by the gauge group and the representation content, with no dependence on the parameters at $M_{Pl}$. Indeed, in the limit $b_G \to \infty$:

$$\alpha_G = \frac{1}{b_G t_G}, \qquad (t_G = \frac{1}{2\pi}\log\frac{M_{Pl}}{M_{GUT}}) \qquad (15)$$

and the value of $Y_t(M_{GUT})$ can be determined by its (quasi)-infra-red fixed point structure at $M_{GUT}$. Using the same eqs.(9-13) but now for the running above the GUT scale and with the coefficients appropriate for the chosen unified model, for $c_G Y_t(M_{Pl})F(M_{GUT}) \gg 1$ we have

$$Y_t(M_{GUT}) \sim Y_t^{QF}(M_{GUT}) = \frac{E(M_{GUT})}{c_G F(M_{GUT})} = Y_t^F(M_{GUT})\frac{1}{1 - \left[\frac{\alpha_G(M_{GUT})}{\alpha_G(M_{Pl})}\right]^{1+\frac{b_{tG}}{b_G}}} \qquad (16)$$

---

[††]In this case the bottom pole mass may be significantly different from the running mass due to the supersymmetric loop corrections [77, 79]. Similar corrections may be even important for the Kobayashi-Maskawa mixing angles [80]

[‡‡]For large $\tan\beta$ this conclusion may be less strong.

In spite of a short range of the running given by $t_G \sim \mathcal{O}(1)$, for large values of $b_G$ and with $Y_t(M_{Pl}) \sim \mathcal{O}(1)$ we indeed get the result of eq.(16). One should also stress that the values $Y_t^{QF}(M_{GUT})$ remains large and therefore $Y_t(M_Z)$ is still close to its QF point in the MSSM.

Infra-red fixed point structure for the Yukawa coupling in the MSSM (or in GUTs) has important implications for the running of the soft SUSY breaking terms[89] and for the electroweak breaking.[90] In brief, some ratios of the soft SUSY breaking terms are similarly driven to their infra-red fixed point values,[45] e.g.

$$\frac{A_t}{M} = const, \qquad \frac{m_Q^2 + m_U^2 + m_{H_2}^2}{M} = const \qquad (17)$$

and become independent of the initial conditions (here $A_t$, $M$, $m$ are the trilinear soft term, gaugino mass and scalar masses, respectively). Effectively, this reduces the number of free parameters of the model. Radiative electroweak breaking with the top quark Yukawa coupling close to its IR fixed point value has been studied in ref.[45, 79, 48].

Although the relevance of the infra-red fixed point for the low energy physics still remains a speculation, the idea is interesting enough to be worth further investigations. An attempt has been recently made [91] to predict quark and squark masses from the infra-red fixed point structure in models with horizontal family symmetry [92].

### 3.3 ORIGIN AND PATTERN OF SOFT TERMS

The origin and pattern of soft supersymmetry breaking terms is closely related to the mechanism of supersymmetry breaking which remains a fundamental open problem. At present, the most attractive mechanism seems to be the one in which supersymmetry breaking is transmitted from the hidden sector to the observable sector by the dilaton and moduli fields of the effective supergravity (see e.g.[58, 93] and references therein. ) Here (as in the rest of this talk) we concentrate mainly on phenomenological aspects of the problem. Recently, the pattern of soft SUSY breaking terms has been vigorously discussed in the context of radiative electroweak symmetry breaking and of flavour non–conservation in flavour changing neutral current (FCNC) processes.

The observed suppression of FCNC transitions is nicely explained in the Standard Model by GIM mechanism. At the same time it is a very strong constraint on physics beyond the SM which, in general, may provide new mechanism for FCNC transitions. Supersymmetric extensions of the SM do indeed contain additional contributions to FCNC transitions from sfermion exchange in loop diagrams.[96, 85] Such effects are generically suppressed only as $\mathcal{O}\left(\frac{M_Z}{M_{\tilde{f}}}\right)$ where $M_{\tilde{f}}$ is a typical sfermion mass matrix entry. They can be potentially dangerous for FCNC transitions if, in the basis in which the fermion mass matrices are diagonal the sfermion mass matrices have large flavour off–diagonal entries. The simplest way out is to assume that the soft scalar masses are flavour diagonal and universal in an electroweak basis at a large scale [4], as indeed happens in the dilaton–dominated supersymmetry breaking scenario in supergravity (see e.g. [94] an references therein). This assumption has be widely used in phenomenological study of the so–called minimal supergravity model i.e. the MSSM with radiative electroweak breaking and initial (universal) conditions for the soft terms fixed at the GUT scale [90]. As mentioned earlier this model is strongly constrained (very few parameters). As a reflection of a sizable degree of fine–tuning, particularly in the very low and very large $\tan\beta$ regions, it gives strong correlations in the low energy spectrum. Another important result is the gaugino–like composition of the lightest neutralino which follows from $\mu \gg m_0$, $M_{1/2}$, as required for proper radiative breaking (the lightest neutralino is an interesting dark matter candidate).

However, it has been recently emphasized that the universality ansatz at the GUT scale seems too simplistic. At the first place, it may be more natural to make assumptions about soft terms at the Planck scale rather then at $M_{GUT}$ (although as long as we do not have a sound theory of supersymmetry breaking one cannot exclude other possibilities, e.g. much lower scales). In this case the running from $M_{Pl}$ to $M_{GUT}$ in simple GUT models generically leads to some well defined non-universality pattern for the soft terms at $M_{GUT}$.[83] An important aspect of this approach is its predictivity. Secondly, in string inspired supergravity models with moduli contributing to supersymmetry breaking the soft terms are flavour dependent (see e.g.[58, 93] and references therein). Unfortunately, it is very difficult to get for them realistic preditions.

Moreover, on the phenomenological side it has been shown that, at least for the SUSY–SO(10) model with the top –bottom–tau Yukawa unification (i.e. large $\tan\beta$) universal initial conditions at $M_{GUT}$ are incompatible with constraints from $b \to s\gamma$ decay and with the cosmological bound $\Omega h^2 < 1$ for the lightest stable particle (neutralino).[48, 84]

Phenomenological study of models with radiative electroweak breaking and non-universal initial conditions has been conducted in a number of papers [82, 83, 48]. There are two important results. Firstly, one needs much less fine-tuning, particularly in the most interesting regions of very low and very large $\tan\beta$. Even more importantly, qualitatively new solutions are obtained with $\mu \ll M_{1/2}$ i.e. with higgsino – like lightest neutralino and chargino. They appear for certain pattern of non-universality [48] which happens to be the only one consistent with the mentioned above constraints. It is interesting that this phenomenologically desired pattern of non-universality follows [97] from the combination of the RG running from $M_{Pl}$ to $M_{GUT}$ in GUT models, with universal initial values at $M_{Pl}$, and of contributions given by the D–term of broken U(1) in models like SO(10), with the reduction of the rank of the gauge group by one at $M_{GUT}$ [98]. The new class of solutions opens interesting options for dark matter [99] and it coincides with the best fit to the electroweak data (Section2)!

Very little can at the moment be said about implications for models with radiative breaking of more general non-universal pattern potentially possible in supergravity models but here, anyway, the main emphasize should first be on the problem of FCNC to which we now return.

The course of the discussion of the FCNC transitions clearly depends on whether we accept universality ansatz (i.e. say, the dilaton induced SUSY breaking) or not. In the first case, at the electroweak scale we obtain the flavour dependent effects in the sfermion sector only of the order of the Cabibbo– Kobayashi–Maskawa mixing [100], consistently with observations. Recently, very interesting progress has been achieved in studying such "minimal" FCNC effects in the framework of GUTs [95]. The universlity ansatz is applied at the Planck scale. The Grand Unified structure of the theory in the range from $M_{Pl}$ to $M_{GUT}$ together with the large top quark Yukawa coupling leads then to very interesting concrete low energy predictions, in particular in the lepton sector. Their verification is within experimental reach.

If we abandon the universality ansatz (and after all the pattern of the quark masses suggests some new family dependent interactions), we face the problem of explaining in another way the observed at low energy approximate simultaneous diagonality, in the same basis, of the fermion and sfermion mass matrices. So, the first thing here is not to predict but to explain this fermion–sfermion mass allignment. Suppose we start with sfermion mass matrices at the large (GUT, Planck) scale such that diagonal and flavour off–diagonal entries are of the same order of magnitude. For squarks there is the possibility to wash out the flavour non–universality by large flavour blind (strong interaction) renormalization effects in the running from the scale where supersymmetry is broken to the low energy scale[101]. Such effects, however, require large ratio of the gaugino to squarks masses at the GUT scale. The problem is particularly severe for sleptons (flavour blind renormalization is now electroweak) and one needs then the gaugino dominated soft SUSY breaking (as in e.g. no–scale models [102]) to confront the experimental constraints[101]. Of course, we are then back to universal (vanishing!) initial scalar masses at the GUT scale, although with different theoretical motivation (gaugino dominance is incompatible with dilaton breaking).

The most straightforward conclusion is that the low energy FCNC data suggest either universal initial conditions – but why? (dilaton dominated supergravity? no–scale models?–the motivation is not yet convincing enough) or some conspiracy (symmetry?[103]) between fermion and sfermion mass generation which goes beyod the conventional supergravity models e.g. in the direction of including family interactions (or maybe even inspires to look for new mechanisms of supersymmetry breaking). Several ideas in this direction have been recently proposed such as dynamical allignement of the fermion and sfermion mass matrices at the electroweak scale [104], the relevance of the infra-red fixed point structure in models with horizontal U(1) gauge symmetry invoked to explain fermion masses [91] and supergravity models with such U(1) symmetry [105]. It is fair to say that those (and other) ideas need further exploration.

# 4 SUMMARY

Important quantitative progress has been achieved in study of the MSSM both in its low energy and weak scale – GUT scale connection aspects. The most plausible parameter region has been identified, with concrete predictions for the superpartner spectrum and for the pattern of soft terms. The latter can be naturally accommodated in some GUT models but its deeper understanding is still missing. The problem of supersymmetry breaking and the origin and pattern of soft terms is no longer merely a theoretical problem but has its very constrained phenomenological level.

I am grateful to all my collaborators for very enjoyable interactions.